\documentclass[pra,twocolumn,showpacs,longbibliography,superscriptaddress]{revtex4-1}
\usepackage{amsmath,amscd,amssymb,color}
\usepackage{graphicx,amsfonts,epsf}
\usepackage{epstopdf}
\usepackage{float}
\usepackage{enumerate}

\newcommand{\ket}[1]{\vert #1 \rangle }
\usepackage{xcolor}
\definecolor{maroon}{RGB}{100,20,20}
\definecolor{dblue}{RGB}{20,20,100}
\usepackage[colorlinks=true,linkcolor=dblue,citecolor=maroon,urlcolor=maroon]{hyperref}
\begin{document}
\title{Classification and measurement of multipartite
entanglement by reconstruction of correlation tensors on an NMR
quantum processor}
\author{Vaishali Gulati}
\email{vaishali@iisermohali.ac.in}
\affiliation{Department of Physical Sciences, Indian
Institute of Science Education \& 
Research Mohali, Sector 81 SAS Nagar, 
Manauli PO 140306 Punjab India.}
\author{Arvind}
\email{arvind@iisermohali.ac.in}
\affiliation{Department of Physical Sciences, Indian
Institute of Science Education \& 
Research Mohali, Sector 81 SAS Nagar, 
Manauli PO 140306 Punjab India.}
\affiliation{Vice Chancellor, Punjabi University Patiala,
147002, Punjab, India}
\author{Kavita Dorai}
\email{kavita@iisermohali.ac.in}
\affiliation{Department of Physical Sciences, Indian
Institute of Science Education \& 
Research Mohali, Sector 81 SAS Nagar, 
Manauli PO 140306 Punjab India.}
\begin{abstract}
We introduce a protocol to classify three-qubit pure states into different
entanglement classes and implement it on an NMR quantum processor. The protocol
is designed in such a way that the experiments performed to classify the states
can also measure the amount of entanglement present in the state.  The
classification requires the experimental reconstruction of the correlation
matrices using 13 operators. The rank of the correlation matrices provide the
criteria to classify the state in one of the five classes, namely, separable,
biseparable (of three types), and genuinely entangled (of two types, GHZ and W).
To quantify the entanglement, a concurrence function is defined which measures
the global entanglement present in the state, using the same 13 operators.
Global entanglement is zero for separable states and non-zero otherwise.  We
demonstrate the  efficacy of the protocol by implementing it on states chosen
from each of the six inequivalent (under stochastic local operations and
classical communication) classes for three qubits. We also implement the
protocol on states picked at random from the state space of three-qubit
pure states.
\end{abstract} 
\maketitle 
\section{Introduction}
\label{intro}
Quantum entanglement is  an invaluable resource for quantum
computing  and its generation, characterization, detection,
and protection have been the subject of several
investigations~\cite{horodecki-rmp-2009}.  The task of
detecting and certifying the presence of entanglement is 
computationally hard and several methods have been
evolved to tackle this problem~\cite{guhne-physrep-2009}.
The discovery of states that exhibit genuine multipartite
entanglement without multipartite correlations has been the
subject of recent debate~\cite{eltschka-quantum-2020}.  A
review of different methods of identifying genuinely
maximally entangled states in a composite quantum system  is
found in~\cite{enriquez-jphys-2016}.  Specifically for three
qubits, two inequivalent classes of maximally entangled
states have been identified, namely the GHZ class and W
class, which are not interconvertible via stochastic local
operations and classical communication
(SLOCC)~\cite{dur-pra-2000} and various aspects of
tripartite entanglement have been reviewed~\cite{cunha-universe-2019}.  
Several studies have
focused on developing efficient protocols to detect the
presence of genuine multipartite entanglement.  An effective
way to detect genuine multipartite entanglement was proposed
based on multipartite concurrence~\cite{li-pra-2015}.  Lower
bounds on detecting genuine tripartite entanglement were
obtained based on positive partial
transposition~\cite{li-scirep-2017}.  Genuine tripartite
entanglement was detected using quantum Fisher
information~\cite{yang-qip-2020}.  A statistical approach to
characterize multipartite entanglement based on moments of
randomly measured correlation functions was demonstrated on
three qubits~\cite{ketterer-quantum-2020}.

Experimental implementations of entanglement detection
schemes were realized on different quantum architectures.
Schemes to prepare a canonical form for general three-qubit
states, and to show the equivalence of a W superposition
state to the GHZ state, were demonstrated
using NMR~\cite{dogra-pra-2015-1,dogra-pra-2015-2}.
Classification of entanglement in arbitrary
three-qubit pure states was performed on an NMR quantum
processor using a set of minimal measurements
~\cite{singh-pra-2018,singh-qip-2018}.  An embedding quantum
simulator was implemented using NMR and used
to study the entangling dynamics of two- and three-qubit
systems~\cite{xin-pra-2018}.  Nonlocal correlations were
detected using a local measurement-based hierarchy using
three NMR qubits~\cite{singh-epjd-2020}.  Three-photon GHZ
and W states were experimentally demonstrated using
entangled photons~\cite{bouwmeester-prl-1999,zhu-pra-2019}.
A cavity QED scheme was proposed to generate $n$-qubit W
states~\cite{zang-opticsexpress-2016}.  Superconducting
phase qubits were used to fully characterize
three-qubit maximally entangled
states~\cite{neeley-nature-2010}.  Evidence of
genuine multipartite three-qubit entanglement was
demonstrated using a single-neutron
interferometer~\cite{erdosi-njp-2013}.  

A general framework was formulated to detect genuine
multipartite entanglement in systems of arbitrary dimensions
based on correlation tensors~\cite{vicente-pra-2011}.  The
ranks of coefficient matrices were taken into account while
classifying the entanglement of arbitrary multipartite pure
states~\cite{wang-pra-2013}.  
Correlation matrices
in the Bloch representation of density matrices were used to
propose new separability criteria for bipartite and
multipartite quantum
states~\cite{zhao-qip-2020}.  The norm of
correlation vectors was used to detect genuine multipartite
entanglement in tripartite quantum
systems~\cite{li-pra-2017,knips-npj-2020}.  A recent work
proposed a family of multipartite separability criteria
based on a correlation tensor which is linear in the density
operator~\cite{sarbicki-pra-2020}.

In this work, we propose a protocol which is experimentally feasible and uses
fewer resources, with the added advantage that the same set of experiments can
be used to classify a state into different entanglement classes as well as to
measure the amount of entanglement present in it.  We classify a random
three-qubit pure state into one of five different SLOCC entanglement classes.
Our protocol successfully classifies the state into genuinely entangled,
biseparable or separable classes, using a  few experimentally obtained
expectation values.  These same expectation values are also used to quantify the
amount of entanglement present in the state without the need of performing
additional experiments.  Our protocol uses 13 expectation values to reconstruct
correlation tensors. The matricization of the correlation tensor provides three
correlation matrices. The ranks of these matrices for a given state classify it
into one of the five entanglement classes.  For quantification of entanglement,
a concurrence function is defined as the sum of squares of the same expectation
values used to construct the correlation matrices. This function is used to
measure global entanglement in the states,  and is zero for separable states and
non-zero otherwise.  It should be noted that we will be working with deviation
density matrices here, as the NMR signal comes from a small fraction of spins.
In this sense the states we are dealing with are `pseudo
entangled'~\cite{oliveira-book,soares-pinto-phil-2012}.

The rest of this paper is organized as follows:~Section~\ref{theory} briefly
describes the theoretical conditions to detect multipartite entanglement using
correlation tensors.  Sections~\ref{tensor} and \ref{condition} discuss the
tensor matricization and the criteria for tripartite entanglement detection and
quantification, respectively.  Section~\ref{results} describes the experimental
results of tripartite entanglement detection on a three-qubit NMR quantum
processor. The details of circuits and NMR pulse sequences to generate various
three-qubit states and estimate their fidelity are given in
Section~\ref{results1}, while Section~\ref{results2} describes the protocol to
compute correlation tensors from experimental NMR observables and contains the
experimental results of the entanglement detection protocol on random states of
all the classes. The results of experimentally measuring the concurrence are
also provided.  Section~\ref{concl} offers some concluding remarks.

\section{Tripartite entanglement conditions 
and ranks of correlation tensors}
\label{theory}
An $n$-partite pure quantum state $\vert \Psi \rangle \in 
H = H_1\otimes \cdots \otimes H_n$ 
is said to be fully separable if it
can be written as a
tensor product of states for every
subsystem~\cite{werner-pra-1989}:
\begin{equation}
\vert {\Psi}\rangle \langle{\Psi} \vert = \vert {\psi_1}\rangle
\langle{\psi_1}\vert \otimes . . . \otimes \vert {\psi_n}\rangle
\langle{\psi_n}\vert, \quad
\vert \psi_i \rangle \in H_i.
\end{equation}
Further, an $n$-partite  pure quantum state
is termed biseparable if it can be written as:
\begin{equation}
|{\Psi}\rangle \langle{\Psi}| = |{\psi_A}\rangle
\langle{\psi_A}|  \otimes |{\psi_{\bar{A}}}\rangle
\langle{\psi_{\bar{A}}}|
\end{equation}
where $A$ denotes a set of subsystems and $\bar{A}$
denotes the remaining subsystems. 
A state that is not separable or biseparable is called  
a genuinely $n$-party entangled state. In this work, we
focus on characterizing entanglement in a three-qubit system.
\subsection{Correlation tensors and matricization}
\label{tensor}
Consider a three-qubit quantum
state $\rho$ in the
Hilbert space ${\cal H}={\cal H}_1^2 \otimes {\cal H}_2^2
\otimes {\cal H}_3^2$ where ${\cal H}^2$ denotes 
the 2-dimensional 
single qubit Hilbert space. Let $\lambda_i,\, i=1,2,3$ denote the generators
of the unitary group SU(2), which together 
with $\lambda_0=I$ ($I$ being
a $2 \times 2$ identity matrix), comprise
an orthogonal basis of Hermitian operators. 
Any state $\rho$ can be decomposed as:
\begin{eqnarray}
	&&\rho = 
	\frac{1}{8}\left[
		I \otimes I \otimes I \right.\nonumber \\
&&+ 
\sum t_{i}^{1} \lambda_i \otimes I \otimes I +
\sum t_{j}^{2} I \otimes \lambda_j \otimes I +  
\sum t_{k}^{3} I \otimes I \otimes \lambda_k 
\nonumber \\
&& + 
\sum t_{ij}^{12} \lambda_i \otimes \lambda_j \otimes I +
\sum t_{ik}^{13} \lambda_i \otimes I \otimes \lambda_k +
\sum t_{jk}^{23} I \otimes \lambda_j \lambda_k  
	 \nonumber \\
&& \left. +  
\sum t_{ijk}^{123} 
	\lambda_i \otimes \lambda_j \otimes \lambda_k \right]
\end{eqnarray}
with $\rho$ being completely characterized by the expectation
values: 
$t_{i}^{1} = {\rm tr}(\rho \lambda_i \otimes I \otimes I)$,
$t_{j}^{2} = {\rm tr}(\rho I \otimes \lambda_j \otimes I)$,
$t_{k}^{3} = {\rm tr}(\rho I \otimes I \otimes \lambda_k)$,
$t_{ij}^{12} = {\rm tr}(\rho \lambda_i \otimes \lambda_j \otimes I)$,
$t_{ik}^{13} = {\rm tr}(\rho \lambda_i \otimes I \otimes \lambda_k)$,
$t_{jk}^{23} = {\rm tr}(\rho I \otimes \lambda_j \otimes \lambda_k)$,
$t_{ijk}^{123} = {\rm tr}(\rho \lambda_i \otimes \lambda_j
\otimes \lambda_k)$.
The expectation values
$
t_{i}^{1},
t_{j}^{2},
t_{k}^{3}
$
are components of tensors of rank one denoted  by 
$T^{(1)}, T^{(2)}, T^{(3)}$, 
$t_{ij}^{12}, t_{ik}^{13}, t_{jk}^{23}$ 
are components of tensors of rank two denoted by        
$T^{(12)}, T^{(13)}, T^{(23)}$,
and $t_{ijk}^{123}$ are components 
of a rank three tensor $T^{123}$.
$T^{(qp)}$ are the two-qubit correlation tensors and 
$T^{(lmn)}$ is the three-qubit correlation tensor.

Matricization of an $n$-qubit correlation tensor $T^{(ij..n)}$
is defined as the process of ``matrix unfolding'' of the tensor, which
leads to a matrix
$T_{\underline{ij}...n}$,
with underlined indices 
joined together to give the column indices and
the remaining (non-underlined) indices being the row
indices. As an example, consider the $(\underline{12},3)$ 
matricization leading to the matrix
$T_{\underline{12}3}$:
\begin{equation}
T_{\underline{12}3} = 
\left(
\begin{array}{cccc}
T_{11k} & T_{11k} & .. & T_{11k} \\
T_{12k} & T_{12k} & .. & T_{12k} \\
... & ... & .... & ... \\
T_{1n_2k} & ... & .... & ... \\
... & ... & .... & ... \\
T_{21k} & ... & .... & ... \\
... & ... & .... & ... \\
T_{2n_2k} & ... & .... & ... \\
... & ... & .... & ... \\
T_{n_1 n_2 k} & ... & .... & T_{n_1 n_2 k} \\
\end{array}
\right)
\label{eqn4}
\end{equation}
where $T_{xyk} =(T_{xy1}...T_{xyn}), k=1,2...n$
is a row vector. 

The matrix $T_{\underline{12}3}$ can be re-written in terms
of the expectation values
$t_{ijk}^{123}$ as:

\begin{equation}
T_{\underline{12}3} = 
\left(
\begin{array}{cccc}
t_{11k}^{123} & t_{11k}^{123} & .. & t_{11k}^{123} \\
t_{12k}^{123} & t_{12k}^{123} & .. & t_{12k}^{123} \\
... & ... & .... & ... \\
t_{1n_2k}^{123} & ... & .... & ... \\
... & ... & .... & ... \\
t_{21k}^{123} & ... & .... & ... \\
... & ... & .... & ... \\
t_{2n_2k}^{123} & ... & .... & ... \\
... & ... & .... & ... \\
t_{n_1 n_2 k}^{123} & ... & .... & t_{n_1 n_2 k}^{123} \\
\end{array}
\right)
\label{eqn5}
\end{equation}

In Dirac notation the matrix can be
written as~\cite{KoBa09}:
\begin{equation}
T_{\underline{12}3} = \sum_{1,2,3} T_{123} \vert 12
\rangle \langle 3 \vert
\end{equation}

\subsection{Entanglement conditions derived from ranks of correlation
tensors}
\label{condition}
Any three-qubit pure state can be written in the 
generalized Schmidt form~\cite{acin-prl-2000}:
\begin{equation}
\vert \psi\rangle = 
a_0 \vert 000\rangle + a_1 e^{i \theta} \vert 100\rangle 
+ a_2 \vert 101\rangle +
a_3 \vert 110 \rangle+ a_4 \vert 111\rangle 
\end{equation}
where $a_i \geq 0$, $ \sum a^2_i = 1$ and $\theta \in [0, \pi]$.  To check how
the states in the canonical basis can  be classified as separable, biseparable, or
genuinely entangled, let $T_{\underline{1}23}, T_{\underline{2}13}$ and
$T_{\underline{3}12} $ be the matrices constructed  with entries of the tensor
$T_{123}$, as defined in the previous subsection.  For a given state, the matrix
ranks of these three matrices determine the entanglement class the state belongs
to.  The three ranks obtained can thus be used to classify all three-qubit pure
states into five classes namely, genuinely entangled (denoted as `GE'),
biseparable (of three types denoted as: `BS-1', `BS-2', `BS-3'), and separable
(denoted as `SEP'). Although the states belonging to the W class always have
rank 3 for all three matrices, states belonging to the GHZ class could have
either rank 2 or rank 3 for all the matrices.  Hence, it is not always possible
to distinguish between GHZ and W entanglement classes using this method, and we
denote such states as belonging to the genuinely entangled (GE) class of
states.  The ranks of the correlation matrices and corresponding entanglement
class category are given in Table~\ref{tab:classification}.

\begin{table}[h]
\caption{Ranks of the correlation matrices and 
the corresponding entanglement class of three-qubit pure states.}
\centering
\begin{tabular}{rp{5mm}l}
\hline
\hline
 Rank of Correlation Matrices && Class\\
\hline
\hline
$T_{\underline{1}23}=T_{\underline{2}31}=T_{\underline{3}12}=2$
or $3$ 
&& Genuinely entangled\\
$T_{\underline{2}31}=T_{\underline{3}12}=3$ ,
$T_{\underline{1}23}=1$ &&
Biseparable-1\\
         $T_{\underline{1}23}=T_{\underline{3}12}=3$ , $T_{\underline{2}31}=1$ 
&&
Biseparable-2\\
         $T_{\underline{1}23}=T_{\underline{2}31}=3$ ,
$T_{\underline{3}12}=1$ 
&&
Biseparable-3\\  $T_{\underline{1}23}=T_{\underline{2}31}=T_{\underline{3}12}=1$ 
&& Separable\\
	 \hline
     \hline
     \end{tabular}
    \label{tab:classification}
\end{table}

The concurrence $C(\psi)$ for a bipartite pure state is given by $C(\psi) =
\sqrt{2(1- Tr \rho ^2 )}$, where $\rho$ is the density
operator corresponding to the reduced state of one of the
systems. A multipartite system  can be
divided up in many ways and thus for a multipartite pure
state 
we can define a
set of concurrences   
$C_j(\psi)= \sqrt{ 2(1 - Tr \rho ^2_j )}$, where $\rho_j$
is a reduced density matrix of the $j$-th
qubit~\cite{coffman-pra-2000,hill-prl-1997}.  As proved in Reference
\cite{guo2021violation}, the total concurrence which is an entanglement measure
(called global entanglement), is related to the expectation values as sums of
squares of them.
For three qubits, the non-negative 
total concurrence is given by:
\begin{equation}
C^2 _T(\psi) = C^2 _1(\psi) + C^2_2(\psi) + C^2_3(\psi) 
\label{eqn7}
\end{equation}
where
$C_1(\psi) = \sqrt{2(1 - {\rm Tr} \rho ^2_1)}$,
$C_2(\psi) = \sqrt{2(1 - {\rm Tr} \rho ^2_2)}$, and 
$C_3(\psi) = \sqrt{2(1 - {\rm Tr} \rho ^2_3)}$ are
concurrences defined for different partitions of the three
qubit system.  

In terms of expectation values,
\begin{eqnarray}
C_T^2(\psi) =&&
\langle XZX \rangle ^2+ \langle XXZ \rangle ^2 + \langle XXX \rangle ^2+
\langle XYY \rangle^2 \nonumber \\
&&+ \langle XZZ \rangle ^2 +\langle YZY \rangle ^2 + \langle YYZ
\rangle ^2+ \langle YXY \rangle ^2 \nonumber \\
&&+ \langle YYX \rangle^2+ \langle YZZ \rangle
^2 +\langle ZXX \rangle ^2 + \langle ZXY \rangle^2 \nonumber \\
&&+ \langle ZXZ \rangle ^2+
\langle ZYX \rangle^2 + \langle ZYY \rangle ^2 
+\langle ZYZ \rangle ^2 \nonumber \\
&&+ \langle ZZX \rangle ^2+ \langle ZZY \rangle ^2+ 
\langle ZZZ \rangle^2 - 1
\end{eqnarray}
where as an illustration, $\langle XZX \rangle$ is the
expectation value of the operator $\sigma_x \otimes \sigma_z \otimes 
\sigma_x$, where $X,Y,Z$ denote the corresponding Pauli
operators $\sigma_x,\sigma_y,\sigma_z$, respectively.

The global entanglement 
$Q(\vert \psi_i \rangle)$ 
is related to the total concurrence 
$C_T(\psi)$ as $C^2 _T(\psi) =  3 Q(|\psi_i\rangle)$~\cite{meyer-jmp-2002,Brennen-qic-2003}. 
$Q(\vert \psi_i \rangle)$ 
is zero only for separable states,  
it is non-zero for
biseparable and entangled states  and satisfies the following properties: 
(i) $ 0 \leq Q(|\psi_i\rangle) \leq 1$  and (ii) $Q(|\psi_i\rangle)$ is invariant under
local unitaries $U_j$.  
$Q(\vert \psi_i \rangle)$ 
has a maximum value of 1 for the GHZ state,
biseparable states are upper bounded by a 
$Q(\vert \psi_i \rangle)$ 
value of $2/3$, while the W state
is upper bounded by  a 
$Q(\vert \psi_i \rangle)$ 
value of $8/9$.  
\section{Experimental reconstruction of correlation tensors}
\label{results}
\subsection{Constructing states of three NMR qubits}
\label{results1}
\begin{figure}[h]
\centering
\includegraphics[scale=1]{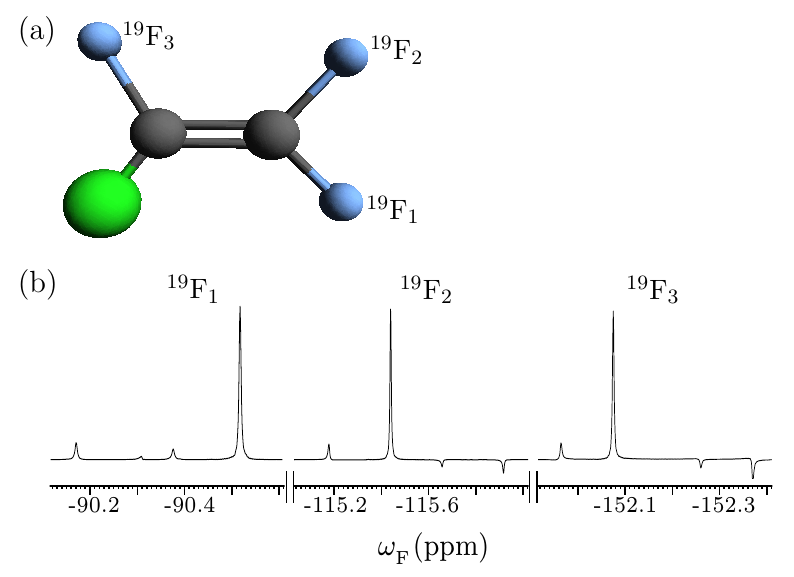}
\caption{(a) Structure of the molecule, trifluoroiodoethylene, 
used to realize the three NMR qubits. (b) NMR spectrum of 
the pseudopure state of the three  ${}^{19}$F qubits.}
\label{fig2-mol}
\end{figure}

The three ${}^{19}$F nuclei in the molecule trifluoroiodoethylene dissolved in
d6-acetone were used to physically realize the three NMR qubits.  
The experimentally determined  T$_1$ and T$_2$ relaxation times for the three
qubits on the average range between 1-5 sec, respectively.  The 
molecular structure and the NMR spectrum of
the PPS state are given in Figure~\ref{fig2-mol}.  
All
experiments were performed at room temperature ($\approx 298$ K) on a Bruker
AVANCE-III 400 MHz NMR spectrometer equipped with a BBO probe.  The NMR
Hamiltonian in the high-temperature, high-field approximation (and assuming a
weak scalar coupling $J_{ij}$ between the spins $i,j$) 
is given by~\cite{oliveira-book}:
\begin{equation}
{\cal H} = - \sum_{i=1}^{3} \omega_i I_{iz}
+ 2 \pi \sum_{i<j}^{3} J_{ij} I_{iz} I_{jz}
\end{equation}
where $\omega_i$ is the chemical shift of the $i$th spin.  The experimentally
measured scalar couplings are given by J$_{12}$= 69.65 Hz, J$_{13}$= 47.67 Hz
and J$_{23}$= -128.32 Hz. The system was initialized in the pseudopure (PPS)
state using the spatial averaging technique~\cite{cory-1998,avikmitra}.  The
density operator of the PPS state is given by:
\begin{equation}
\rho_{000} = \frac{(1- \epsilon)}{8}\mathbb{I}_8 +\epsilon
\vert 000 \rangle \langle 000\vert
\end{equation}
where $\epsilon \sim 10^{-5}$ is the spin polarization at
room temperature and $\mathbb{I}_8$ is the $8 \times 8$
identity operator. The identity part of the density operator plays no role 
and the NMR signal arises only from the contribution of the second
part of the above equation.

The rf pulses for 
pseudopure state preparation~\cite{singh-pra-2018-3udd} were designed using the
Gradient Ascent Pulse Engineering (GRAPE) technique~\cite{Khaneja-jmr-2005}. 
The duration of the single-qubit gates was around 
600 $\mu$s, while for two-qubit gates, 
the durations of the pulses were set to be around 1/2J, where
J is the strength of the scalar coupling between the two relevant qubits. 
The system was evolved from PPS to other
states via state-to-state transfer unitaries, with pulse durations of
$\approx 20$ms,  and average state fidelities of $\geq 0.99$.

\begin{figure}[h]
\centering
\includegraphics[scale=1]{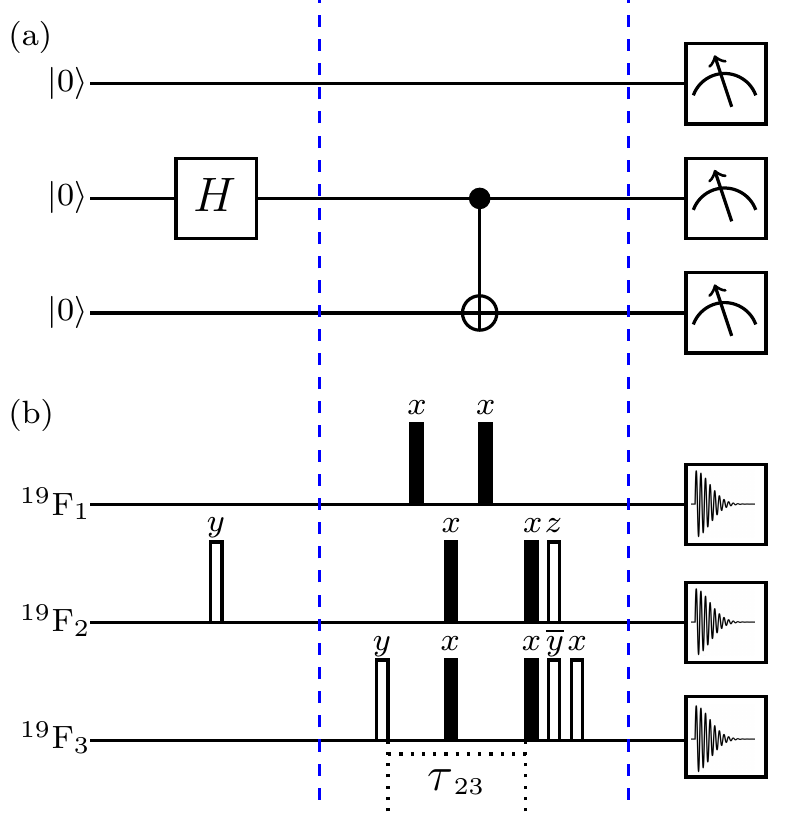}
\caption{{(a) Quantum circuit showing the sequence of
implementation of the single-qubit and two-qubit controlled
gates required to construct the BS-1 state  and (b)
the corresponding NMR pulse sequence for the experimental
implementation of  the BS-1 state.  The broad filled
rectangles denote $\pi$ pulses and the unfilled rectangles
denote $\pi/2$ pulses; the phases of the pulses  are written
above each pulse. The time interval 
$\tau_{23}$  is set to $1/2 J_{F_{2}F_{3}}$.}}
\label{fig3-ckt}
\end{figure}

We experimentally prepared states from each of the six inequivalent (under
SLOCC) classes of three-qubit states, namely the GHZ state, the W state, three
biseparable states (denoted as `BS-1', `BS-2', and `BS-3', respectively) and a
separable state (denoted as `SEP'), using single-qubit unitary rotations and
two-qubit CNOT gates. 
We chose to use the $\vert 111 \rangle$ state as an example of an SEP state and
it was prepared by applying a single-qubit rotation of $\pi/2$ on all the three
qubits in the initial PPS state.  The BS-1 state was prepared by applying an rf
pulse inducing a $(-\pi/2)$ rotation with $y$ phase on the second qubit followed
by a CNOT$_{23}$ gate (details of the circuit and corresponding NMR pulse
sequence are given in Figure~\ref{fig3-ckt}).  The BS-2  state was prepared by
applying a Hadamard gate on the first qubit followed by a CNOT$_{13}$ gate.
Similarly, the BS-3 state was prepared by applying a Hadamard gate on the first
qubit followed by a CNOT$_{12}$ gate and then a $\pi$ rotation of $x$ phase on
the first qubit.  

\begin{figure}[h]
\centering
\includegraphics[scale=1]{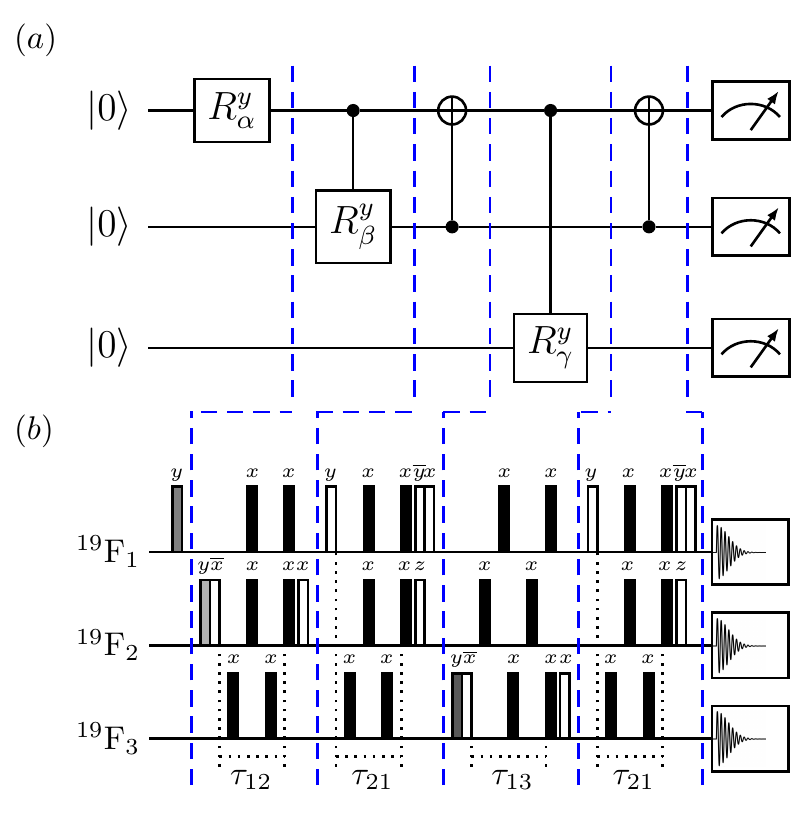}
\caption{(a) Quantum circuit showing the sequence of implementation of the
single-qubit and two-qubit controlled gates required to construct the W state
and (b) the corresponding NMR pulse sequence 
for the experimental implementation of
the W state.  The broad filled black 
rectangles denote $\pi$ pulses and the unfilled
rectangles denote $\pi/2$ pulses. The three gray rectangles 
denote the $2\alpha, 2\beta $ and $2\gamma $ rotations, where
$\alpha=\frac{\pi}{3}$, $\beta=\arcsin{(\frac{1}{\sqrt{3}})}$ and
$\gamma=\frac{\pi}{4}$, respectively. 
The phases of the pulses  are written above
each pulse. The time interval $\tau_{ij}$  is set to the corresponding  
$1/2 J_{ij}$, where $J_{ij}$ denotes 
the strength of the coupling between the $i,j$ qubits.}
\label{w}
\end{figure}

The quantum circuit for GHZ state preparation can be found in
References~\cite{dogra-pra-2015-1} and \cite{singh-pra-2018}.  The W state
prepared in the canonical basis was constructed with two controlled-rotation
gates and CNOT gates, starting from the PPS state, as shown in Figure~\ref{w}.
The unitaries to be implemented are written inside the boxes representing the
gates.  The angles $\alpha,\beta$ and $\gamma$ were set to
$\alpha=\frac{\pi}{3}$, $\beta=\arcsin{(\frac{1}{\sqrt{3}})}$ and
$\gamma=\frac{\pi}{4}$, respectively.  The rf pulse durations to implement an
angle of $\pi/2$ at power level of 28.59 W is 16.2 $\mu$s.

The standard methods for quantum state reconstruction 
for NMR quantum information processing typically
involve performing 
full state tomography~\cite{long-qst,leskowitz} which is 
computationally expensive, although some alternatives involving
maximum likelihood estimation have been proposed and used
in our group~\cite{singh-pla-2016}.
For this work, we have used a least squares constrained
convex optimization method to reconstruct the density 
matrix of the desired state~\cite{gaikwad-ijqi-2020}.
Fidelities of the experimentally reconstructed
states (as compared to the theoretically expected state)
were computed using the Uhlmann-Jozsa 
measure~\cite{jozsa,uhlmann}:
\begin{equation} 
{\mathcal F}(\chi^{}_{\rm
expt},\chi^{}_{{\rm theo}})= \frac{|{\rm Tr}[\chi^{}_{\rm
expt}\chi_{\rm theo}^\dagger]|} {\sqrt{{\rm Tr}[\chi_{\rm
expt}^\dagger\chi^{}_{\rm expt}] {\rm Tr}[\chi_{\rm
theo}^\dagger\chi^{}_{\rm theo}]}} \label{e12}
\end{equation}
where $\chi^{}_{{\rm theo}}$ and $\chi^{}_{\rm expt}$ denote the theoretical and
experimental density operators, respectively.  We experimentally prepared the
PPS with a fidelity of 0.96$\pm$0.01.  The experimental fidelities for the GHZ
and W states were 0.95$\pm$0.01 and $0.93 \pm 0.02$, 
respectively.  The BS-1, BS-2,
and BS-3 states were prepared with fidelities 0.95$\pm$0.02, 0.95$\pm$0.01, and
0.97$\pm$0.02, respectively, while the fidelity of the SEP state was
0.96$\pm$0.01.
\subsection{Reconstructing correlation tensors from NMR observables}
\label{results2}
The reconstruction of the full correlation tensor 
(as described in Section~\ref{theory}) 
requires experimentally measuring 27
expectation values of the form $\sigma_j \otimes \sigma_l
\otimes \sigma_m$ ($j,l,n=1,2,3$), where $\sigma$ represents
the Pauli matrices. For the sake of simplicity, the expectation
operators are denoted by combinations of  $X,Y,Z$ where $X$
represents the Pauli matrix $\sigma_x$ and so on. When
the correlation tensors are constructed for any state, some of the expectation
values become zero, while others are related to one another.
Out of 27 observables needed to construct a correlation matrix, 8 of them
become zero in the canonical basis: 
\begin{eqnarray}
\langle XXY \rangle &=& \langle XYX \rangle= \langle XYZ \rangle= \langle XZY
\rangle=0 \nonumber \\
\langle YXX \rangle &=& \langle YXZ \rangle= \langle YYY \rangle= 
\langle YZX \rangle=0 \nonumber \\
\end{eqnarray}
Using the following relationships among the expectation values:
\begin{eqnarray}
\langle XXX \rangle &=& -\langle XYY \rangle= -\langle YXY 
\rangle= -\langle YYX \rangle \nonumber \\
\langle XXZ \rangle &=& -\langle YYZ \rangle \nonumber \\
\langle XZX \rangle &=& -\langle YZY \rangle \nonumber \\
\langle ZXY \rangle &=&  \langle ZYX \rangle
\end{eqnarray}
6 of the observables can be 
computed from the others.
This leaves us with the following
13 observables that are required to be experimentally measured:
\begin{eqnarray}
&&\langle XXX \rangle,\,\,
\langle XXZ \rangle, \,\,
\langle XZX \rangle,  \,\,
\langle ZXY \rangle, \nonumber \\
&& \langle XZZ \rangle, \,\,
\langle YZZ \rangle, \,\,
\langle ZXX \rangle,  \,\,
\langle ZXZ \rangle, \nonumber \\
&& \langle ZYY \rangle, \,\,
\langle ZYZ \rangle, \,\,
\langle ZZX \rangle, \,\,
\langle ZZY \rangle, \,\,
\langle ZZZ \rangle
\end{eqnarray}
The correlation matrices are computed from 
13 expectation values, and are constructed according to
Eq.~(\ref{eqn5}).
Here $t_{111}^{123}={\rm tr}(\sigma_1 \otimes \sigma_1 \otimes
\sigma_1)=\langle XXX\rangle,t_{121}^{123}
={\rm tr}(\sigma_1 \otimes \sigma_2 \otimes
\sigma_1)=\langle XYX\rangle$ etc. The three correlation matrices are
given by:

\begin{eqnarray}
T_{\underline{1}23} &=& 
\begin{pmatrix}
t_{111}^{123} &0 &t_{113}^{123} & 0& -t_{111}^{123}& 0 & t_{131}^{123}& 0& t_{133}^{123} \\
0 &-t_{111}^{123} &0 & -t_{111}^{123} & 0 & -t_{113}^{123} & 0& -t_{131}^{123}& t_{233}^{123} \\
t_{311}^{123} &t_{312}^{123} &t_{313}^{123} & t_{312}^{123}& t_{322}^{123}& t_{323}^{123} & t_{331}^{123} & t_{332}^{123} & t_{333}^{123} 
\end{pmatrix}
\nonumber \\
T_{\underline{2}13} &=&
\begin{pmatrix}
t_{111}^{123} &0 &t_{113}^{123} & 0& -t_{111}^{123}& 0 & t_{311}^{123}& t_{312}^{123}& t_{313}^{123} \\

0 &-t_{111}^{123} &0 & -t_{111}^{123} & 0 & -t_{113}^{123} & t_{312}^{123}& t_{322}^{123}& t_{323}^{123} \\

t_{131}^{123} &0 &t_{133}^{123} & 0& -t_{131}^{123}& t_{233}^{123} & t_{331}^{123} & t_{332}^{123} & t_{333}^{123} 
\end{pmatrix}
\nonumber \\
T_{\underline{3}12} &=& 
\begin{pmatrix}
t_{111}^{123} &0 &t_{131}^{123} & 0& -t_{111}^{123}& 0 & t_{311}^{123}& t_{312}^{123}& t_{331}^{123} \\

0 &-t_{111}^{123} &0 & -t_{111}^{123} & 0 & -t_{131}^{123} & t_{312}^{123}& t_{322}^{123}& t_{332}^{123} \\
t_{113}^{123} &0 &t_{133}^{123} & 0& -t_{113}^{123}& t_{233}^{123} & t_{313}^{123} & t_{323}^{123} & t_{333}^{123} 
\end{pmatrix}
\nonumber \\
\end{eqnarray}

The ranks of these three correlation matrices are then calculated and
the entanglement class is verified according to Table
\ref{tab:classification}.  The final step involves
quantifying the entanglement by calculating total
concurrence given in Eq.~(\ref{eqn7}), 
using the same experimentally measured set of 13
expectation values. 

\begin{figure}[h] 
\centering 
\includegraphics[scale=1.0]{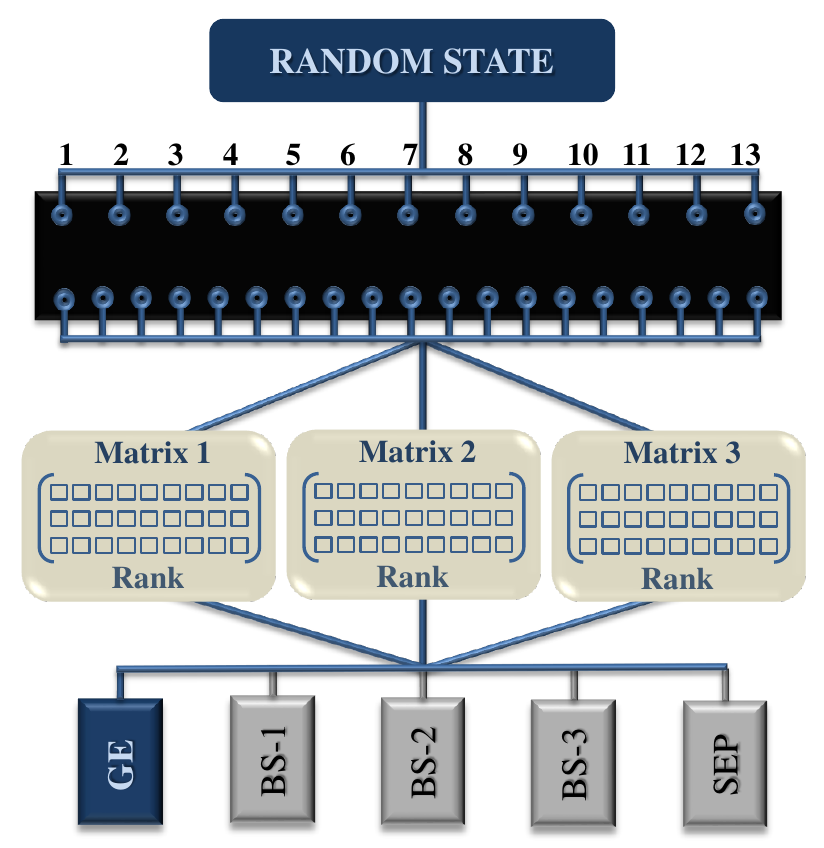}
\caption{Schematic of the entanglement classification protocol.  
The blue box at the top represents an unknown pure three-qubit state. 
The inputs to the black box
represent experimental data for the 13 expectation values. 
The outputs from the black
box denote all 
the expectation values that are required to construct 
the correlation matrices. All
three correlation matrices required for every state make use of these
expectation values. The rank of the three correlation matrices 
are computed to 
classify the state as belonging to one of the
five entanglement classes: Genuinely Entangled (GE), 
Biseparable (BS-1,BS-2,BS-3), Separable (SEP).
} 
\label{fig1-diagram} 
\end{figure}

The schematics of the entanglement classification protocol
is shown in Figure~\ref{fig1-diagram}. The protocol begins
with a random state as an input.  The black box represents
experimental reconstruction of the correlation matrix.  The
13 observables are calculated which are shown as inputs to
the black box, and the output consists of all the
observables not experimentally measured but needed to
construct the correlation matrix.  
The rank of the three correlation
matrices obtained for each state are calculated and
classified according to Table~\ref{tab:classification}. 

\begin{table}[h]
\caption{Ranks and entanglement class 
characterization of three-qubit experimental states.}
\vspace*{12pt}
\centering
\begin{tabular}{|c|c|c|c|c|}
\hline
\hline
State &	$R(T_{\underline{1}23})$ &  $R(T_{\underline{2}31})$ &
$R(T_{\underline{3}12})$  & Class\\
\hline
\hline
GHZ&2&2&2&GE\\
W&3&3&3&GE\\
BS-1&1&3&3&BS-1\\
BS-2&3&1&3&BS-2\\
BS-3&3&3&1&BS-3\\
SEP&1&1&1&SEP\\
R$_1$&1&1&1&SEP\\
R$_2$&1&3&3&BS-1\\
R$_3$&3&1&3&BS-2\\
R$_4$&1&1&1&SEP\\
R$_5$&1&3&3&BS-1\\
R$_6$&1&1&1&SEP\\
R$_7$&3&1&3&BS-2\\
R$_8$&3&3&3&GE\\
R$_9$ &3&3&3&GE\\
R$_{10}$&1&3&3&BS-1\\
R$_{11}$&3&1&3&BS-2\\
R$_{12}$&3&3&1&BS-3\\
R$_{13}$&3&1&3&BS-2\\
R$_{14}$&3&3&3&GE\\
R$_{15}$&3&3&3&GE\\
R$_{16}$&3&3&1&BS-3\\
R$_{17}$&3&3&3&GE\\
R$_{18}$&3&1&3&BS-2\\
R$_{19}$&3&3&3&GE\\
R$_{20}$&3&3&3&GE\\
\hline
\hline
\end{tabular}
\label{tab:rank}
\end{table}

The states that we experimentally constructed 
from each SLOCC class are:
\begin{eqnarray*}
{\rm GHZ} &=& \frac{1}{\sqrt{2}}(\ket{000}+\ket{111}) \\
{\rm W}   &=& \frac{1}{\sqrt{4}}(\ket{000}+\ket{100}+\ket{101}+\ket{110})\\
{\rm BS-}1 &=& \frac{1}{\sqrt{2}}(\ket{000}+\ket{011})\\
{\rm BS-}2 &=& \frac{1}{\sqrt{2}}(\ket{000}+\ket{101})\\
{\rm BS-}3 &=& \frac{1}{\sqrt{2}}(\ket{010}+\ket{100})\\
{\rm SEP} &=&\ket{111}
\end{eqnarray*}

\begin{figure}[ht] 
\centering 
\includegraphics[scale=1.0]{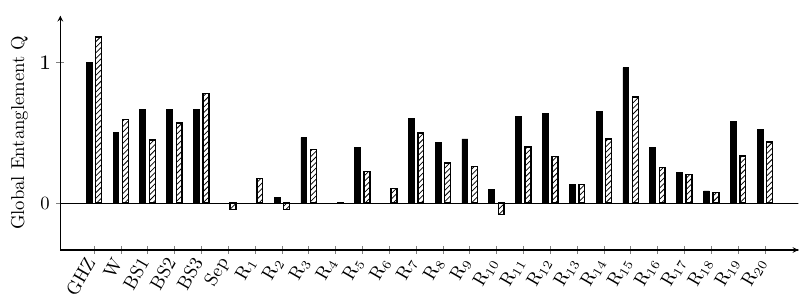}
\caption{Bar plots of the theoretically expected and the experimentally computed
global entanglement $Q(\vert \psi_i \rangle)$ 
of three-qubit pure states.  The horizontal axis labels the
state, while the vertical axis represent the global
entanglement values.  Black and
cross-hatched bars represent the theoretical and experimental values,
respectively.} 
\label{fig5} 
\end{figure}

The pure states of three qubits were prepared in the canonical basis, which
includes the six known states from each SLOCC class as well as arbitrarily
prepared random states.  The random states were generated using the  Mathematica
package and the state preparation circuits to prepare the states experimentally
were designed using  the Mathematica package UniversalQCompiler~\cite{math}.
The pulse circuits thus obtained were implemented experimentally and the states
obtained had fidelities in the range 80 to 97$\%$.  
The experimentally prepared states had several errors due to
experimental noise, errors in pulse calibration parameters etc.  These errors
lead to inaccuracies in the computed expectation values since they are
calculated from the experimental matrices.  The experimental matrices themselves
do not produce exactly zero values of matrix elements where there is zero
expected theoretically.  Further, since we are finding the rank of the matrix,
(the number of independent rows or columns) it is important to be aware of the
numerical values. In the known matrices with good fidelity, these ``zero
errors'' were checked and it was found that the values can go up to 0.09. So the
error bar was fixed for these values in all the random states in the range
$0\pm 0.09$, below which value they can be considered to be zero.  The
expectation values were calculated from the matrices and were used to construct
the correlation matrices labeled $T_{\underline{1}23}$, $T_{\underline{2}31}$,
$T_{\underline{3}12}$. The rank of the computed matrices were matched with the
theoretical values (Table~\ref{tab:classification}). The ranks of experimentally
prepared GHZ, W, BS-1, BS-2, BS-3 and SEP states after error correction match
well with the ranks in Table~\ref{tab:classification}, proving the efficacy of
the protocol. The ranks of the twenty randomly generated states labeled
(R$_1$,R$_2$,R$_3$ etc.. ) were used to classify the states into one of the five
entanglement classes (Table~\ref{tab:rank}). The protocol was able to
correctly characterize the entanglement class of all the twenty randomly
generated states, with no ambiguity.

It is useful in some situations 
to quantify the amount of entanglement present in the state, and
several quantities exist to quantify entanglement.
We use concurrence as a measure of entanglement, which is
able to quantify entanglement not only in genuinely
entangled states but also in biseparable states. 
We use the same expectation
values that we used to classify the states, to measure 
the entanglement present in them. 
The global entanglement $Q(\vert \psi_i \rangle)$ 
for known states was calculated for experimentally obtained
states and compared with the theoretically expected values. 
The results are depicted in
Figure~\ref{fig5},
where theoretical and experimental results 
of the computed global entanglement are compared
for known states as well as for the randomly generated states. 
The values of the global entanglement match well
with the theoretically expected values,
within experimental errors. 
\section{Conclusions} 
\label{concl}
We designed and experimentally implemented a protocol to classify the
entanglement class of and to measure the entanglement in random three-qubit pure
states.  We reconstructed the correlation tensors experimentally by measuring
only a few observables. The ranks of the subsequently computed correlation
matrices provide the criteria for classification of multipartite entanglement.
The protocol works well for pure states and was experimentally demonstrated on
states belonging to the six inequivalent (under SLOCC) classes for three qubits
as well as on twenty randomly generated states.  The same expectation values
that were used to characterize entanglement class of the state,  were used to
measure the amount of global entanglement present in the state under
consideration.

To be able to correctly characterize, certify and detect multipartite
entanglement in multiqubit systems using minimal experimental resources is a
challenging task and our work is a step forward in this direction.  Future work
would involve extending the experimental protocol for implementation on larger
qubit registers and for mixed states. 

\begin{acknowledgments}
All experiments were performed on a 600 MHz FT-NMR
spectrometer at the NMR Research Facility IISER Mohali.
Arvind acknowledges financial support from
DST/ICPS/QuST/Theme-1/2019/General Project number Q-68.
K.D. acknowledges financial support from
DST/ICPS/QuST/Theme-2/2019/General Project number Q-74.
\end{acknowledgments}

%

\end{document}